\def\PM{$\pm$}
\documentstyle[12pt]{article}

\font\bx=cmbx10 at 17.32truept

\title{\large \bx Measure of Autocorrelation Times of Local Hybrid Monte Carlo
Algorithm
for Lattice QCD \\ }

\author{
Paolo Marenzoni \\
Luigi Pugnetti \\
Dipartimento di Fisica\\
Universit\`a di Parma, \\
Viale delle Scienze \\
I-43100 Parma \\
Italy
\\
\\
Pietro Rossi \\
CRS4 \\
Via Nazario Sauro, 10 \\
I-09123 Cagliari \\
Italy
}

\date{}

\begin{document}

\maketitle
\begin{abstract}

\setlength{\baselineskip}{22pt}

\noindent
We report on a study of the autocorrelation times of the local version
of the Hybrid Monte Carlo (LHMC) algorithm for pure gauge $SU(3)$. We
compare LHMC to standard multi-hit Metropolis and to the global
version of the same HMC. For every algorithm we measure the
autocorrelation time ($\tau$) for a variety of observables and the
string tension ($\sigma$) as a function of $\beta$.  The measurements
performed on $8^4$ and $16^4$ lattices indicate that the
autocorrelation time of LHMC is significantly shorter than for the
other two algorithms.

\end{abstract}

\newpage

\setlength{\baselineskip}{20pt}

\centerline{\bx Introduction}

In order to extract a meaningful number from a Monte Carlo (MC)
simulation we have to associate a realistic error estimate to the
measurement of the mean value. The ability of obtaining a good value for
the error depends on our knowledge of the number of statistical independent
samples that we have, which in turn implies a knowledge of the autocorrelation
time $\tau$ associated to the chosen algorithm.

In lattice QCD, as we approach the continuum limit ( $a\to 0$, or $\beta
\to \infty$)
we observe the well known phenomenon of critical slowing down.
This means that, as $a\to 0$, $\tau$ tends to diverge, thus requiring more and
more iterations of a given algorithm to obtain an independent configuration.
The approach to the continuum limit implies that the autocorrelation length
of the system is diverging, or the lowest mass, expressed in lattice units
is diverging. For a given lattice, a numerical simulation will be
meaningful only if we
limit ourselves to values of the coupling constant corresponding to
correlation length significantly shorter than the lattice size, so for
a given lattice the critical slowing down can be described by a critical
exponent $z$ parameterizing the dependence of $\tau$ on  the typical
mass scale, $\tau = \alpha m^{-z}.$\\

In chapter 1 we give a brief overview of the HMC algorithm and the
LHMC version, chapter two contains a description of the measurement of
the autocorrelation time, chapter three describes in details the
measurement performed, including a discussion on the choice of the
optimal trajectory length.  Results are presented and commented in
chapter four.

\section{Hybrid Monte Carlo}

 The HMC algorithm is one of a variety of Monte Carlo techniques to
simulate statistical systems described by some local action
\cite{hmc1}.

 We will concentrate on lattice gauge theories, therefore
 the action we are talking about is the usual Wilson's action,

\begin{equation}
 S_p = \sum_{x,\mu,\nu}\beta[1-\frac{1}{3}Re Tr(U_{\mu}(n)U_{\nu}(n+\mu)
 U_{-\mu}(n+\mu+\nu)U_{-\nu}(n+\nu))].\
\end{equation}

 The HMC prescription dictates to introduce fictitious conjugate momenta
 $\pi$ for every degree of freedom, and simulate the pseudo-Hamiltonian

 \begin{equation}
 H(\pi,U) = Tr \pi^{2}_{\mu}(x) + S_p(U).
 \end{equation}
 The simulation begins by generating the momenta $\pi$'s distributed
 according to the Gaussian:

 \begin{equation}
 P(\pi) = exp\left({\sum_{x,\mu}Tr \pi^{2}_{\mu}(x)}\right),
 \end{equation}
 then, starting from the randomly chosen momenta and the current
 $U$'s a new set of $\pi$'s and $U$'s are chosen as the end point
 of the trajectory generated by integrating Hamilton's equation of motion
 associated to the Hamiltonian (2).

 At the end of the trajectory, the small variation in energy introduced
 by the finite step size is corrected by a Metropolis-like step.

 All the details on this algorithm can be found in \cite{hmc1}
 \cite{piehmc}. The key point to observe is that, since all the degrees
 of freedom are evolved simultaneously, in order to preserve detailed balance,
 the Metropolis step must be applied to the whole lattice.
 This step consists in comparing the difference between the initial
 and the final value  of the energy
 \begin{equation}
  \delta H = H_{new} - H_{old}
 \end{equation}
 and the probability of accepting the new
 configuration is proportional to
 \begin{equation}
 P\approx e^{-\delta H}.
 \end{equation}
 $\delta H $ is an extensive quantity and it will grow with some
 power of the volume. In turn, this implies that, in order to keep
 the acceptance constant, the step size will have to be reduced
 according to the same volume's law.

\subsection{Local Hybrid Monte Carlo}
 Let us denote with $U_E$ the gauge field variables sitting at even sites of
 the lattice and $U_O$ those sitting at odd sites. The stochastic matrix
 describing the updating from the old variables $(U_E, U_O)$ to the
 new variables $(\overline{U_E},\overline{U_O})$ will be denoted by:
\begin{equation}
    W(\overline{U_E},\overline{U_O};U_E, U_O).
\end{equation}
 If we choose to update even sites first, then odd sites,
 this stochastic matrix factorizes in
\begin{equation}
    W(\overline{U_E},\overline{U_O};\overline{U_E}, U_O)
    W(\overline{U_E},U_O; U_E, U_O).
\end{equation}
Since the Wilson action only involves nearest neighbor interactions,
even sites are totally independent, so are odd ones and, as a result, each
 one of the factors of the above equation factorizes. For instance:
\begin{equation}
 W(\overline{U_E},U_O; U_E, U_O) = \prod_{x\epsilon E} W(\overline{U_x},U_O;
U_x, U_O).
\end{equation}

 Parallelism can be achieved easily by applying simultaneously all the
elemental
 stochastic matrices to one checkerboard at a time. For each variable then, we
 apply the HMC scheme and the final Metropolis acceptance-rejection test
 only involves a local modification of the action therefore there is not
 any degrading of the acceptance due to the volume and we can use very large
step sizes.

 The integration of Hamilton's equations of motion involves the computation of
 the force term which is made up by the staple of the link under investigation.
 This staple is constructed solely by links belonging to the opposite
 checkerboard, so the algorithm is equivalent to the integration of
 the equation of motion for a variable in an external field. This is, of
 course, an other clear advantage over the conventional HMC. In the global
 version of HMC, at every step the staple has to be computed anew, therefore
 the computational cost of a trajectory is a multiple of the cost of the
 first step. In the local version only the first step is expensive,
 successive steps are only a fraction of the cost that can be estimated around
10\%.

 As we will see in the section devoted to the discussion of the results, the
combination
 of these effects account for an inexpensive algorithm that moves
 effectively through phase space.

\section{Autocorrelation times}

For a set of $n$ measures of an observable $A$ we can define the
autocorrelation function:
\begin{equation}
 C(t) = N\sum_{i=1}^{n-|t|} <(A_i - <A>)(A_{i+t} - <A>)>
\end{equation}
where $<A>$ is the average over all the measurements and $N$ is a
constant fixed by the condition $C(0)=1$. If we assume that, at
large times $t$, this function decays exponentially $C(t)\approx
exp[-t/\tau_{exp}]$,
then $\tau_{exp}$ provides us with a definition of the
autocorrelation time. This way to compute $\tau$ however
can be affected by large errors, when $\tau$ is small.

This defines, of course, only the autocorrelation of the given
observables. Different ones can in principle, and often do, have
 different autocorrelation times. To define the autocorrelation
 of a given algorithm, one should survey a complete set of observables and
 chose the highest $\tau$. This, being highly non practical, is approximated
by a choice of selected operators and that will, operatively, define
the $\tau$ of the algorithm.

Another definition of $\tau$ is given by \cite{sokal}:
\begin{equation}
 \tau_{int} = \frac{1}{2}\sum_{t=-\infty}^{\infty} C(t)
\end{equation}
However, this estimator of $\tau_{int}$ present a whole new set of troubles
because when the sum over $t$ goes to $|t|>>\tau$ the function $C(t)$ contains
noise much higher than the signal. Adding these terms we would
over estimate the errors. The practical solution is a cut off in the sum in
the previous equation, introducing a suitably constant $M$:
\begin{equation}
 \tau_{int} = \frac{1}{2}\sum_{t=-M}^{M} C(t)
\end{equation}
In this way our calculation will be biased, since we are systematically
neglecting terms in the sum (6).
On the other hand we can systematically exploit that bias defining
the sum extended over $t$'s as long as $C(t) > {1\over e^q}.$
Our choice is $ q = 10.$

 Whatever definition of $\tau$ we choose, we noticed consistently that
 in order to have a reliable estimate, the length of
 our run must be approximately 1000 times the estimated autocorrelation
 time.

 Our $\tau$  have been measured on a large set of Wilson loop,
 from $1 \times 1 $ to $ 3 \times 3.$ For these observables
 the behavior is not noticeably different, therefore the data shown
 in the following refer to the $3 \times 3 $ Wilson loops.

\section{Tuning of the parameters}
 There are two parameters characterizing HMC like algorithms,
 $N_s$ and $\delta t.$ It is well known that for the global HMC
 tuning of these parameters is crucial to improve on the autocorrelation
 of the algorithm. For the LHMC we have noticed that the autocorrelation
 time, of the monitored observables, depends very lightly on the
 parameters chosen. A value of $ T = N_s\delta t = 1 $ is nearly optimal
 in every case and all data shown have been obtained with $T$ between
 $0.9$ and $1.2$, with $N_s$ varying from 2 to 4.
 These values correspond to an acceptance rate around $90\%.$

 When discussing autocorrelation times, these should be correctly given
 in CPU time and not sweeps but for the LHMC, as explained above,
 extra steps have only a marginal cost, so we give, for sake of simplicity,
 all the $\tau$'s expressed in number of iterations.
 The same holds for multi-hit Metropolis, while in the case of global HMC,
 where different trajectories length have a very different cost, we limited
ourselves
 to the fixed trajectory length of $T=0.3$, with $N_s=10$ \cite{Gupta}.
 All the simulations for multi-hit Metropolis were done with 8 hits per
step.

\section{Results}
 Autocorrelation times have been measured, besides for LHMC, for
multi-hit Metropolis and the global HMC.  In table 1 me show the
autocorrelation times for the $ 3 \times 3 $ Wilson loop on a $16^4 $
lattice, for all the three algorithms considered, while tables 2 and 3
concentrate on LHMC for different observables and $\beta$'s.

 From the data it is obvious that LHMC decorrelates much faster when
measured in terms of sweeps, and even better if measured in terms of
computational cost.  This concept is emphasized in table 4 where we
plot the time it takes to perform a number of sweeps corresponding to
one $\tau$.  For instance, at $\beta = 6.0 $, if we assign, in
arbitrary units, 1 to LHMC it takes 7.4 for multi-hit Metropolis and
400 for global HMC.

Our goal was the measure of the dynamical
critical exponent of the algorithm in the hope of confirming certain
theoretical results obtained on the Gaussian model, claiming it to be
one. This was not possible since we noticed an unexpected behavior of
the $\tau$'s.  For every observables that we analyzed, we observed
that, plotting $\tau$ versus the autocorrelation length (or for that
matter versus $\beta $), we did not get a monotonically increasing
function, as expected, but a peak at a finite value of the coupling.
The position of the peak is strongly observable dependent and less
strongly volume dependent. Without a monotonic behavior it is not
possible to fit the $\tau$ to a power law of the correlation length
and therefore extract the dynamical critical exponent of the
algorithm. This unexpected behavior is not an artifact of the
algorithm used, since we observe it for multi-hit metropolis as well
as with global HMC, but rather a feature of the physics of lattice
QCD.  Some author recently, in the framework of a different algorithm
\cite{DeF}, observing the same behavior, conjectured some connection
to the finite temperature phase transition. That interpretation is not
fully consistent with previously well known results, about the
position of that transition. From our analysis it is not possible
today to offer support for that view since it is quite puzzling that
the peak moves about according to the observables.  Furthermore, even
for the larger Wilson loops, the position of the peak is at values of
$\beta$ much lower than the value observed by the above authors.

\subsection*{Acknowledgments}
The authors warmly acknowledge useful discussions with E. Onofri.  All
the numerical works presented in this paper was performed on
Connection Machines CM2 at the University of Parma, CM200 at CRS4 in
Cagliari and CM5 at IPG in Paris. We wish to extend our thanks to the
above mentioned institutions.\\
This research was partly supported by the Sardegna Regional Authority.
\vfill\eject

{\bf \noindent Captions}\\
\\
{\bf \noindent Table 1:}

LHMC and Metropolis $\tau$ for the $3\times 3$ Wilson's loop.\\
\vskip 0.3cm
{\bf \noindent Table 2:}

$\tau$ values for a variety of Wilson's loops.\\
\vskip 0.3cm
{\bf \noindent Table 3:}

Creutz's ratios.\\
\vskip 0.3cm
{\bf \noindent Table 4:}

Ratios of timings to cover one $\tau$.\\

\vfil\eject

\clearpage

\begin{table}
\hfil{\begin{tabular}{|c|c|c|c|}\hline
 $\beta$ &  LHMC         & Metropolis       & HMC\\ \hline
         &                &               & \\ \hline
 5.6     & 3.93 \PM 1.07 & 14.35 $\pm$ 1.91 & 36 $\pm$ 12  \\ \hline
         &                &                &\\ \hline
 5.8     & 8.27 \PM 1.32 & 81.36 $\pm$ 37.20&              \\ \hline
         &                &               & \\ \hline
 6.0     & 5.57 \PM 0.27 & 33.22 $\pm$ 15.99&              \\ \hline
         &                &                &\\ \hline
 6.2     & 5.32 \PM 1.62 & 15.43 $\pm$ 7.25 & 88 $\pm$ 50  \\ \hline
\end{tabular}\hfil}
\caption{}
\end{table}

\begin{table}
\begin{tabular}{|c|c|c|c|c|c|c|}\hline
 $\beta$ &    $w11$       &       $w21$    &     $w22$      &  $w31$         &
 $w32$        &   $w33$        \\ \hline
 5.4    &  8.14 $\pm$ 0.36 &                &                &                &
               &                 \\ \hline
 5.6    &  9.56 $\pm$ 0.16 & 9.95 $\pm$ 0.09 & 9.87 $\pm$ 0.21 &  9.30 $\pm$
0.50 &  8.42 $\pm$ 1.34 &  3.93 $\pm$ 1.07  \\ \hline
 5.7    &                &                &                &                &
             &  7.07 $\pm$ 1.11  \\ \hline
 5.8    &  7.15 $\pm$ 0.26 &  8.59 $\pm$ 0.48 & 9.82 $\pm$ 0.23  &  8.82 $\pm$
0.19 & 10.70 $\pm$ 0.47 &  8.27 $\pm$ 1.32 \\ \hline
 5.9    &                &                &                &                &
             &  7.69 $\pm$ 0.19 \\ \hline
 6.0    &  2.71 $\pm$ 0.12 &  3.65 $\pm$ 0.86 &  4.13 $\pm$ 0.22 &  3.38 $\pm$
0.44 &  5.34 $\pm$ 0.07 &  5.57 $\pm$ 0.27  \\ \hline
 6.2    &  2.36 $\pm$ 0.34 &  2.79 $\pm$ 0.42 &  3.16 $\pm$ 0.45 &  2.74 $\pm$
0.34 &  4.01 $\pm$ 0.40 &  5.32 $\pm$ 1.62    \\ \hline
 6.3    &  2.12 $\pm$ 0.18 &                &                &                &
               &  3.77 $\pm$ 0.29  \\ \hline
 6.43 & 2.58 $\pm$ 0.75  & & & & & \\ \hline
 6.5  & 2.21 $\pm$ 0.39  & & & & & \\ \hline
\end{tabular}
\caption{ }
\end{table}

\newpage

\section*{}

\begin{table}
 \hfil{\begin{tabular}{|c|c|c|} \hline
 $\beta$ &  $\chi_{2,2}$  & $\chi_{3,2}$   \\ \hline
         &                &                \\ \hline
 5.6     & 6.73 \PM 1.16  &  4.19 \PM 0.56 \\ \hline
         &                &                \\ \hline
 5.8     & 6.63 \PM 0.08  &  5.56 \PM 0.61 \\ \hline
         &                &                \\ \hline
 6.0     & 3.09 \PM 0.03  &  3.48 \PM 0.14 \\ \hline
         &                &                \\ \hline
 6.2     & 2.09 \PM 0.14  &  2.81 \PM 0.48 \\ \hline
\end{tabular}}\hfil
\caption{}
\end{table}

\begin{table}
 \hfil{\begin{tabular}{|c|c|c|} \hline
&& \\
$\beta$   & $\displaystyle{\tau_{Metropolis}\over \tau_{LHMC}}$ &
            $\displaystyle{\tau_{HMC}\over \tau_{LHMC}}$ \\
&& \\ \hline
  5.6     &  4.6  & 169 \\ \hline
  5.8     &  12.2 &     \\ \hline
  6.0     &  7.4  & 408 \\ \hline
  6.2     &  3.6  &     \\ \hline
\end{tabular}}\hfil
\caption{ }
\end{table}


\newpage

\begin{thebibliography}{10}

\bibitem{hmc1}
S.~Duane, A.D.~Kennedy, B.J.~Pendleton and D.~Roweth,
\newblock{\bf Phys. Lett. B195} (1987) 216.


\bibitem{piehmc}
K.~Bitar and A.D.~Kennedy, R.~Horsley and S.~Meyer, P.~Rossi,
\newblock{\bf Nucl. Phys. B313} (1989) 377.

\bibitem{sokal}
N.~Madras and A.D.~Sokal,
\newblock{\bf Jour. of Stat. Phys. 50} (1988) 109.

\bibitem{Gupta}
R.~Gupta, G.W.~Kilcup and S.R.~Sharpe,
\newblock{\bf Phys. Rev. D38} (1988) 1278.

\bibitem{DeF}
K.~Akemi et al.
\newblock{\bf IPS Research Report No.92-28, HUPD-9218} (1992).

\end{thebibliography}
\end{document}